\documentstyle[11pt]{article}
\begin{document}

\title{New holographic scalar field models of dark energy in non-flat universe}
\author{K. Karami\thanks{E-mail: KKarami@uok.ac.ir}\\J. Fehri\\\small{Department of
Physics, University of Kurdistan, Pasdaran St., Sanandaj,
Iran}\\
}

\maketitle

\begin{abstract}
Motivated by the work of Granda and Oliveros [L.N. Granda, A.
Oliveros, Phys. Lett. B {\bf 671}, 199 (2009)], we generalize
their work to the non-flat case. We study the correspondence
between the quintessence, tachyon, K-essence and dilaton scalar
field models with the new holographic dark energy model in the
non-flat FRW universe. We reconstruct the potentials and the
dynamics for these scalar field models, which describe accelerated
expansion of the universe. In the limiting case of a flat
universe, i.e. $k = 0$, all results given in [L.N. Granda, A.
Oliveros, Phys. Lett. B {\bf 671}, 199 (2009)] are obtained.
\end{abstract}
\clearpage
\section{Introduction}
Type Ia supernovae observational data suggest that our universe is
experiencing an accelerated expansion driven by an exotic energy
with negative pressure which is so-called dark energy (DE)
\cite{Riess}. However, the nature of DE is still unknown, and
people
 have proposed some candidates to describe it. The cosmological
 constant, $\Lambda$, is the most obvious theoretical candidate of
 DE, which has the equation of state $\omega=-1$.
 Astronomical observations indicate that the cosmological constant is many orders of magnitude
 smaller than estimated in modern theories of elementary particles \cite{Weinberg}. Also the
 "fine-tuning" and the "cosmic coincidence" problems are the two
 well-known difficulties of the cosmological constant problems
 \cite{Copeland}.

 There are different alternative theories for the dynamical DE scenario which have been
 proposed by people to interpret the accelerating universe. i) The scalar field models of DE including
 quintessence \cite{Wetterich},
 phantom (ghost) field \cite{Caldwell1}, K-essence \cite{Chiba}
 based on earlier work of K-inflation \cite{Picon3},
 tachyon field \cite{Sen}, dilatonic ghost condensate \cite{Gasperini},
 quintom \cite{Elizalde1}, and so forth. ii) The interacting DE models including
 Chaplygin gas \cite{Kamenshchik}, braneworld models \cite{Deffayet}, and agegraphic DE models \cite{Cai}, etc.

Recently, a new DE candidate, based on the holographic principle,
was proposed \cite{Horava}. According to the holographic
principle, the number of degrees of freedom in a bounded system
should be finite and has relations with the area of its boundary
\cite{Hooft}. In quantum field theory a short distance (UV)
cut-off $\Lambda$ is related to a long distance (IR) cut-off $L$
due to the limit set by formation of a black hole, which results
in an upper bound on the zero-point energy density \cite{Cohen}.
By applying the holographic principle to cosmology, one can obtain
the upper bound of the entropy contained in the universe
\cite{Fischler}. Following this line, Li \cite{Li} argued that for
a system with size $L$ and UV cut-off $\Lambda$, it is required
that the total energy in a region of size $L$ should not exceed
the mass of a black hole of the same size, thus
$L^3\rho_{\Lambda}\leq LM_P^2$, where $\rho_{\Lambda}$ is the
quantum zero-point energy density caused by UV cut-off $\Lambda$
and $M_P$ is the reduced Planck Mass $M_P^{-2}=8\pi G$. The
largest $L$ allowed is the one saturating this inequality, thus
$\rho_{\Lambda}=3c^2M_P^2L^{-2}$, where $c$ is a numerical
constant. Recent observational data, which have been used to
constrain the holographic DE model, show that for the non-flat
universe $c=0.815_{-0.139}^{+0.179}$ \cite{Li5}, and for the flat
case $c=0.818_{-0.097}^{+0.113}$ \cite {Li6}. Also Li \cite{Li}
showed that the cosmic coincidence problem can be resolved by
inflation in the holographic DE model, provided the minimal number
of e-foldings \cite{Li}. The holographic models of DE have been
studied widely in the literature
\cite{Enqvist,Elizalde2,Guberina}. As we mentioned before, the UV
cut-off is related to the vacuum energy, and IR cut-off is related
to the large scale of the universe, for example Hubble horizon,
future event horizon or particle horizon. Taking $L$ as the size
of the current universe, for instance, the Hubble scale, the
resulting energy density is comparable to the present day DE.
However, as found by Hsu \cite{Hsu}, in that case, the evolution
of the DE is the same as that of dark matter (dust matter), and
therefore it cannot drive the universe to accelerated expansion.
The same appears if one chooses the particle horizon of the
universe as the length scale $L$ \cite{Li}. An interesting
proposal is made by Li \cite{Li}: Choosing the event horizon of
the universe as the length scale, the holographic DE not only
gives the observation value of DE in the universe, but also can
drive the universe to an accelerated expansion phase. In that
case, however, an obvious drawback concerning causality appears in
this proposal. Event horizon is a global concept of space–time;
existence of event horizon of the universe depends on future
evolution of the universe; and event horizon exists only for
universe with forever accelerated expansion. This motivated Granda
and Oliveros \cite {Granda1} to propose a new infrared cut-off for
the holographic DE, which besides the square of the Hubble scale
also contains the time derivative of the Hubble scale. This model
depends on local quantities and avoids the problem of causality
which appears using the event horizon area as the IR cut-off. They
also in an another paper \cite{Granda2} using the new infrared
cut-off for the holographic DE \cite {Granda1}, studied the
correspondence between the quintessence, tachyon, K-essence and
dilaton energy density with this holographic DE density for the
case of DE dominance in the flat FRW universe. This correspondence
allowed them to reconstruct the potentials and the dynamics for
the scalar field models, which describe accelerated expansion.

Besides, as usually believed, an early inflation era leads to a
flat universe. This is not a necessary consequence if the number
of e-foldings is not very large \cite{Huang}. It is still possible
that there is a contribution to the Friedmann equation from the
spatial curvature when studying the late universe, though much
smaller than other energy components according to observations.
Therefore, it is not just of academic interest to study a universe
with a spatial curvature marginally allowed by the inflation model
as well as observations. Some experimental data have implied that
our universe is not a perfectly flat universe and that it
possesses a small positive curvature \cite{Bennett}.

In the light of all mentioned above, it becomes obvious that using
the holographic DE model with the new infrared cut-off introduced
by \cite{Granda1} and extending the work \cite{Granda2} to a
non-flat case is well motivated. To do this, in Section 2, we
obtain the equation of state parameter for the new holographic DE
model given by \cite{Granda1} in the context of the a non-flat
universe. In Section 3, like \cite{Granda2}, we suggest a
correspondence between the holographic and the quintessence,
tachyon, K-essence and dilaton scalar field models in the presence
of a spatial curvature. We reconstruct the potentials and the
dynamics for these scalar field models, which describe accelerated
expansion. Section 4 is devoted to conclusions.

\section{New holographic DE model in non-flat universe}
Following \cite{Granda1}, the holographic DE density with the new
infrared cut-off is given by
\begin{equation}
\rho_{\Lambda}=3M_{P}^{2}(\alpha H^{2}+\beta\dot{H}),\label{rhoh}
\end{equation}
where $H=\dot{a}/a$ is the Hubble parameter and $\alpha$ and $\beta$
are constants which must satisfy the restrictions imposed by the
current observational data.

Here we consider the Friedmann-Robertson-Walker (FRW) universe with
line element
\begin{equation}
ds^{2}=-dt^{2}+a^{2}(t)\Big(\frac{dr^{2}}{1-kr^{2}}+r^{2}d
\Omega^{2}\Big),
\end{equation}
where $k$ denotes the curvature of space $k=0,1,-1$ for flat, closed
and open universe, respectively. A closed universe with a small
positive curvature ($\Omega_k\sim 0.02$) is compatible with
observations \cite{Bennett}. Like \cite{Granda2}, restricting our
study to the current cosmological epoch, we neglect the
contributions from matter and radiation and assume the DE is
dominated in the presence of spatial curvature. Therefore the first
friedmann equation becomes
\begin{equation}
H^{2}+\frac{k}{a^{2}}=\frac{1}{3M_{P}^{2}}\rho_{\Lambda}.\label{F}
\end{equation}
Substituting Eq. (\ref{rhoh}) in (\ref{F}) gives
\begin{equation}
\frac{{\rm d}H^2}{{\rm
d}x}+\frac{2}{\beta}(\alpha-1)H^2=\frac{2}{\beta}ke^{-2x},
\end{equation}
where $x=\ln a$. Integrating the above equation with respect to $x$
yields
\begin{equation}
H^{2}=\frac{k}{\alpha-\beta-1} e^{-2x} +\gamma
e^{-\frac{2}{\beta}(\alpha-1)x},\label{H2-1}
\end{equation}
where $\gamma$ is an integration constant. If we set $k=0$ for the
flat case and using $\dot{x}=H$, then Eq. (\ref{H2-1}) reduces to
\begin{equation}
H^2=\frac{\beta}{\alpha-1}\dot{H},\label{H2-2}
\end{equation}
where dot denotes the time derivative with respect to the cosmic
time $t$. Taking the integral in both sides of Eq. (\ref{H2-2}) with
respect to $t$, we get
\begin{equation}
H=\frac{\beta}{\alpha-1}\frac{1}{t},\label{Hflat}
\end{equation}
which is same as Eq. (2.2) in \cite{Granda2}. From the conservation
equation
\begin{equation}
\dot{\rho_{\Lambda}}+3H(1+\omega_{\Lambda})\rho_{\Lambda}=0,
\end{equation}
and using Eq. (\ref{rhoh}), the equation of state (EoS) parameter,
$\omega_{\Lambda}=p_{\Lambda}/\rho_{\Lambda}$, can be obtained as
\begin{equation}
\omega_{\Lambda}=-1-\frac{2\alpha H \dot{H}+\beta\ddot{H}}{3H(\alpha
H^{2}+\beta \dot{H})}.\label{wh1}
\end{equation}
Replacing $H$ from Eq. (\ref{H2-1}) in (\ref{wh1}) yields
\begin{equation}
\omega_{\Lambda}=-\frac{1}{3}\left(\frac{k\Big(\frac{\alpha-\beta}{\alpha-\beta-1}\Big)e^{-2x}+\gamma\Big(\frac{3\beta-2\alpha+2}{\beta}\Big)e^{-\frac{2}{\beta}(\alpha-1)x}}
{k\Big(\frac{\alpha-\beta}{\alpha-\beta-1}\Big)e^{-2x}+\gamma
e^{-\frac{2}{\beta}(\alpha-1)x}}\right),\label{wh2}
\end{equation}
which is a time-dependent EoS parameter. When the EoS parameter of
DE is time-dependent, then it can transit from
$\omega_{\Lambda}>-1$ to $\omega_{\Lambda}<-1$ \cite{Wang}. The
analysis of the properties of DE from recent observations mildly
favor models with $\omega_{\Lambda}$ crossing -1 in the near past
\cite{Alam}.

If we set $k=0$, we obtain $\omega_{\Lambda}$ for the flat
universe as
\begin{equation}
\omega_{\Lambda}=-1+\frac{2}{3}\frac{\alpha-1}{\beta},\label{wflat}
\end{equation}
which is same as Eq. (2.5) in \cite{Granda2}. Contrary to the
non-flat case, the EoS parameter for the flat case is constant.
However, Granda and Oliveros \cite{Granda2} showed that by
applying some restrictions on the constants $\alpha$ and $\beta$,
the EoS parameter of the new holographic DE in the flat case, Eq.
(\ref{wflat}), can behave as a quintessence or a phantom type DE.
But, neither the quintessence nor the phantom alone can fulfill
the transition from $\omega_{\Lambda}>-1$ to $\omega_{\Lambda}<-1$
and vice versa.
\section{Correspondence between the new holographic and scalar field models of DE}
Here like \cite{Granda2}, we suggest a correspondence between the
new holographic DE model with the quintessence, tachyon, K-essence
and dilaton scalar field models but in the context of the non-flat
universe. To establish this correspondence, we compare the new
holographic energy density (\ref{rhoh}) with the corresponding
scalar field model density and also equate the equations of state
for this models with the EoS parameter given by (\ref{wh2}).

\subsection{New holographic quintessence model}

The energy density and pressure of the quintessence scalar field
$\phi$ are as follows \cite{Copeland}
\begin{equation}
\rho_{Q}=\frac{1}{2}\dot{\phi}^{2}+V(\phi),\label{rhoq}
\end{equation}
\begin{equation}
p_{Q}=\frac{1}{2}\dot{\phi}^{2}-V(\phi).
\end{equation}
The EoS parameter for the quintessence scalar field is given by
\begin{equation}
\omega_{Q}=\frac{p_{Q}}{\rho_{Q}}=\frac{\dot{\phi^{2}}-2V(\phi)}{\dot{\phi^{2}}+2V(\phi)}.\label{wq}
\end{equation}
If we establish the correspondence between the new holographic DE
scenario and the quintessence DE model, then using Eqs. (\ref{wh2})
and (\ref{wq}) we have
\begin{equation}
\omega_{\Lambda}=-\frac{1}{3}\left(\frac{k\Big(\frac{\alpha-\beta}{\alpha-\beta-1}\Big)e^{-2x}+\gamma\Big(\frac{3\beta-2\alpha+2}{\beta}\Big)e^{-\frac{2}{\beta}(\alpha-1)x}}
{k\Big(\frac{\alpha-\beta}{\alpha-\beta-1}\Big)e^{-2x}+\gamma
e^{-\frac{2}{\beta}(\alpha-1)x}}\right)=\frac{\dot{\phi^{2}}-2V(\phi)}{\dot{\phi^{2}}+2V(\phi)},
\end{equation}
also using Eqs. (\ref{rhoh}) and (\ref{rhoq}) we get
\begin{equation}
\rho_{\Lambda}=3M_{p}^{2}(\alpha H^{2}+\beta
\dot{H})=\frac{1}{2}\dot{\phi}^{2}+V(\phi),
\end{equation}
then one can obtain the kinetic energy term and the quintessence
potential energy as follows
\begin{equation}
\dot{\phi}^2=2M_{P}^2\Big[k\Big(\frac{\alpha-\beta}{\alpha-\beta-1}\Big)
e^{-2x}+\gamma\Big(\frac{\alpha-1}{\beta}\Big)
e^{-\frac{2}{\beta}(\alpha-1)x}\Big],\label{phi2q}
\end{equation}
\begin{equation}
V(\phi)=M_{P}^{2}\Big[2k\Big(\frac{\alpha-\beta}{\alpha-\beta-1}\Big)
e^{-2x}+\gamma\Big(\frac{3\beta-\alpha+1}{\beta}\Big)
e^{-\frac{2}{\beta}(\alpha-1)x}\Big].\label{vphiq}
\end{equation}
From Eqs. (\ref{H2-1}) and (\ref{phi2q}), using $\dot{\phi}=\phi'H$
where prime denotes the derivative with respect to $x$, we obtain
\begin{equation}
\phi'=\sqrt{2}M_{P}\left[\frac{k\Big(\frac{\alpha-\beta}{\alpha-\beta-1}\Big)
e^{-2x}+\gamma\Big(\frac{\alpha-1}{\beta}\Big)
e^{-\frac{2}{\beta}(\alpha-1)x}}{\frac{k}{\alpha-\beta-1} e^{-2x}
+\gamma e^{-\frac{2}{\beta}(\alpha-1)x}}\right]^{1/2}.
\end{equation}
Consequently, after integration with respect to $x$ we can obtain
the evolutionary form of the quintessence scalar field as
\begin{equation}
\phi(a)-\phi(0)=\sqrt{2}M_{P}\int_{0}^{\ln
a}\left[\frac{k\Big(\frac{\alpha-\beta}{\alpha-\beta-1}\Big)
e^{-2x}+\gamma\Big(\frac{\alpha-1}{\beta}\Big)
e^{-\frac{2}{\beta}(\alpha-1)x}}{\frac{k}{\alpha-\beta-1} e^{-2x}
+\gamma e^{-\frac{2}{\beta}(\alpha-1)x}}\right]^{1/2}{\rm
d}x,\label{phi3q}
\end{equation}
where we take $a_0=1$ for the present time.

For the flat case if we set $k=0$, using Eqs. (\ref{H2-1}) and
(\ref{Hflat}), assuming $\phi(0)=0$ for the present time $t_0=0$,
then Eqs. (\ref{vphiq}) and (\ref{phi3q}) reduce to
\begin{equation}
\phi(t)=\sqrt{\frac{2\beta}{\alpha-1}}~M_P\ln t,
\end{equation}
\begin{equation}
V(\phi)=\beta\left[\frac{3\beta-\alpha+1}{(\alpha-1)^2}\right]
M_P^2\exp\left({-\sqrt{\frac{2(\alpha-1)}{\beta}}\frac{\phi}{M_P}}\right),
\end{equation}
which are same as Eqs. (3.5) and (3.6) in \cite{Granda2},
respectively. Note that in Eq. (3.6) in \cite{Granda2}, a factor
$\beta$ has been missed.

\subsection{New holographic tachyon model}
The tachyon field was proposed as a source of the dark energy. The
tachyon is an unstable field which has become important in string
theory through its role in the Dirac-Born–-Infeld (DBI) action
which is used to describe the D-brane action \cite{Sen}. The
effective Lagrangian density of tachyon matter is given by
\cite{Sen}
\begin{equation}
{\mathcal{L}}=-V(\phi)\sqrt{1+\partial_{\mu}\phi
\partial^{\mu}\phi}.
\end{equation}
The energy density and pressure for the tachyon field are as
following \cite{Sen}
\begin{equation}
\rho_{T}=\frac{V(\phi)}{\sqrt{1-\dot{\phi}^{2}}},\label{rhot}
\end{equation}
\begin{equation}
p_{T}=-V(\phi)\sqrt{1-\dot{\phi}^{2}},
\end{equation}
where $V(\phi)$ is the tachyon potential. The EoS parameter for
the tachyon scalar field is obtained as
\begin{equation}
\omega_{T}=\frac{p_{T}}{\rho_{T}}=\dot{\phi}^{2}-1.\label{wt}
\end{equation}
If we establish the correspondence between the new holographic DE
and tachyon DE, then using Eqs. (\ref{wh2}) and (\ref{wt}) we have
\begin{equation}
\omega_{\Lambda}=-\frac{1}{3}\left(\frac{k\Big(\frac{\alpha-\beta}{\alpha-\beta-1}\Big)e^{-2x}+
\gamma\Big(\frac{3\beta-2\alpha+2}{\beta}\Big)e^{-\frac{2}{\beta}(\alpha-1)x}}
{k\Big(\frac{\alpha-\beta}{\alpha-\beta-1}\Big)e^{-2x}+\gamma
e^{-\frac{2}{\beta}(\alpha-1)x}}\right)=\dot{\phi}^{2}-1,\label{htw}
\end{equation}
also comparing Eqs. (\ref{rhoh}) and (\ref{rhot}) one can write
\begin{equation}
\rho_{\Lambda}=3M_{P}^{2}(\alpha H^{2}+\beta
\dot{H})=\frac{V(\phi)}{\sqrt{1-\dot{\phi}^{2}}}.\label{htd}
\end{equation}
From Eqs. (\ref{htw}) and (\ref{htd}), one can obtain the kinetic
energy term and the tachyon potential energy as follows
\begin{equation}
\dot{\phi}^2=\frac{2}{3}\left[\frac{k\Big(\frac{\alpha-\beta}{\alpha-\beta-1}\Big)
e^{-2x}+\gamma\Big(\frac{\alpha-1}{\beta}\Big)
e^{-\frac{2}{\beta}(\alpha-1)x}}{k\Big(\frac{\alpha-\beta}{\alpha-\beta-1}\Big)
e^{-2x}+\gamma e^{-\frac{2}{\beta}(\alpha-1)x}}\right],\label{phi2t}
\end{equation}
\begin{eqnarray}
V(\phi)=\sqrt{3}M_{P}^{2}\left\{\Big[k\Big(\frac{\alpha-\beta}{\alpha-\beta-1}\Big)
e^{-2x}+\gamma\Big(\frac{3\beta-2\alpha+2}{\beta}\Big)
e^{-\frac{2}{\beta}(\alpha-1)x}\Big]\right.\nonumber\\\left.\times\Big[k\Big(\frac{\alpha-\beta}{\alpha-\beta-1}\Big)e^{-2x}+\gamma
e^{-\frac{2}{\beta}(\alpha-1)x}\Big]\right\}^{1/2}.\label{vphit}
\end{eqnarray}
From Eqs. (\ref{H2-1}) and (\ref{phi2t}), using
$\dot{\phi}=\phi'H$, we obtain the evolutionary form of the
tachyon scalar field as
\begin{equation}
\phi(a)-\phi(0)=\sqrt{\frac{2}{3}}\int_{0}^{\ln
a}\left(\frac{k\Big(\frac{\alpha-\beta}{\alpha-\beta-1}\Big)
e^{-2x}+\gamma\Big(\frac{\alpha-1}{\beta}\Big)
e^{-\frac{2}{\beta}(\alpha-1)x}}{\Big[k\Big(\frac{\alpha-\beta}{\alpha-\beta-1}\Big)
e^{-2x}+\gamma
e^{-\frac{2}{\beta}(\alpha-1)x}\Big]\Big[\frac{k}{\alpha-\beta-1}e^{-2x}
+\gamma e^{-\frac{2}{\beta}(\alpha-1)x}\Big]}\right)^{1/2}{\rm
d}x.\label{phi3t}
\end{equation}
For the flat case if we set $k=0$, using Eqs. (\ref{H2-1}),
(\ref{Hflat}), assuming $\phi(0)=0$ for the present time $t_0=0$,
then Eqs. (\ref{vphit}) and (\ref{phi3t}) reduce to
\begin{equation}
\phi(t)=\sqrt{\frac{2(\alpha-1)}{3\beta}}~t,
\end{equation}
\begin{equation}
V(\phi)=2M_P^2\left(1-\frac{2(\alpha-1)}{3\beta}\right)^{1/2}\left(\frac{\beta}{\alpha-1}\right)\frac{1}{\phi^2},
\end{equation}
which are same as Eqs. (3.11) and (3.12) in \cite{Granda2},
respectively.
\subsection{New holographic K-essence model}

The K-essence scalar field model of DE is also used to explain the
observed late-time acceleration of the universe. The K-essence is
described by a general scalar field action which is a function of
$\phi$ and $\chi=\dot{\phi}^2/2$, and is given by \cite{Chiba,
Picon3}
\begin{equation}
S=\int {\rm d} ^{4}x\sqrt{-{\rm g}}~p(\phi,\chi),
\end{equation}
where $p(\phi,\chi)$ corresponds to a pressure density as
\begin{equation}
p(\phi,\chi)=f(\phi)(-\chi+\chi^{2}),
\end{equation}
and the energy density of the field $\phi$ is
\begin{equation}
\rho(\phi,\chi)=f(\phi)(-\chi+3\chi^{2}).\label{rhok}
\end{equation}
The EoS parameter for the K-essence scalar field is obtained as
\begin{equation}
\omega_{K}=\frac{p(\phi,\chi)}{\rho(\phi,\chi)}=\frac{\chi-1}{3\chi-1}.\label{wk}
\end{equation}
Equating Eq. (\ref{wk}) with the new holographic EoS parameter
(\ref{wh2}), $\omega_{K}=\omega_{\Lambda}$, we find the solution
for $\chi$
\begin{equation}
\chi=\frac{1}{3}\left(\frac{2k\Big(\frac{\alpha-\beta}{\alpha-\beta-1}\Big)e^{-2x}+\gamma\Big(\frac{3\beta-\alpha+1}{\beta}\Big)e^{-\frac{2}{\beta}(\alpha-1)x}}
{k\Big(\frac{\alpha-\beta}{\alpha-\beta-1}\Big)e^{-2x}+
\gamma\Big(\frac{2\beta-\alpha+1}{\beta}\Big)e^{-\frac{2}{\beta}(\alpha-1)x}}\right).\label{xi1k}
\end{equation}
Using Eq. (\ref{xi1k}), $\dot{\phi}^2=2\chi$, and
$\dot{\phi}=\phi'H$, we obtain the evolutionary form of the
K-essence scalar field as
\begin{equation}
\phi(a)-\phi(0)=\sqrt{\frac{2}{3}}\int_{0}^{\ln
a}\frac{1}{H}\left(\frac{2k\Big(\frac{\alpha-\beta}{\alpha-\beta-1}\Big)e^{-2x}+\gamma\Big(\frac{3\beta-\alpha+1}{\beta}\Big)e^{-\frac{2}{\beta}(\alpha-1)x}}
{k\Big(\frac{\alpha-\beta}{\alpha-\beta-1}\Big)e^{-2x}+\gamma
\Big(\frac{2\beta-\alpha+1}{\beta}\Big)e^{-\frac{2}{\beta}
(\alpha-1)x}}\right)^{1/2}{\rm d}x,\label{phik}
\end{equation}
where $H$ is given by Eq. (\ref{H2-1}).

Comparing Eqs. (\ref{rhoh}) and (\ref{rhok}),
$\rho_{\Lambda}=\rho(\phi,\chi)$, using Eqs. (\ref{F}) and
(\ref{H2-1}), one can obtain a relation for $f(\phi)$ as follows
\begin{equation}
f(\phi)=\frac{3M_P^2(1-3\omega_{\Lambda})^2}{2(1-\omega_{\Lambda})}\left
[k\left(\frac{\alpha-\beta}{\alpha-\beta-1}\right)e^{-2x}+\gamma
e^{-\frac{2}{\beta}(\alpha-1)x}\right],\label{fphi}
\end{equation}
where $\omega_{\Lambda}$ is given by Eq. (\ref{wh2}).

For the flat case if we set $k=0$, using Eqs. (\ref{Hflat}),
(\ref{wflat}), assuming $\phi(0)=0$ for the present time $t_0=0$,
then Eqs. (\ref{xi1k}), (\ref{phik}) and (\ref{fphi}) reduce to
\begin{equation}
\chi=\frac{1}{3}\left(\frac{3\beta-\alpha+1}{2\beta-\alpha+1}\right),\label{xi2k}
\end{equation}
\begin{equation}
\phi(t)=\left[\frac{2}{3}\left(\frac{3\beta-\alpha+1}{2\beta-\alpha+1}\right)\right]^{1/2}t,
\end{equation}
\begin{equation}
f(\phi)=6M_{P}^2\beta\left[\frac{2\beta-\alpha+1}{(\alpha-1)^2}\right]\frac{1}{\phi^2},
\end{equation}
which are same as Eqs. (3.18), (3.19) and (3.20) in \cite{Granda2},
respectively.
\subsection{New holographic dilaton field}

The dilaton scalar field model of DE is obtained from the low-energy
limit of string theory. It is described by a general
four-dimensional effective low-energy string action. The coefficient
of the kinematic term of the dilaton can be negative in the Einstein
frame, which means that the dilaton behaves as a phantom-type scalar
field. However, in presence of higher-order derivative terms for the
dilaton field $\phi$ the stability of the system is satisfied even
when the coefficient of $\dot{\phi}^2$ is negative \cite{Gasperini}.
The pressure (Lagrangian) density and the energy density of the
dilaton DE model is given by \cite{Gasperini}
\begin{equation}
p_{D}=-\chi+c'e^{\lambda\phi}\chi^{2},
\end{equation}
\begin{equation}
\rho_{D}=-\chi+3c'e^{\lambda\phi}\chi^{2},\label{rhod}
\end{equation}
where $c'$ and $\lambda$ are positive constants and
$\chi=\dot{\phi}^2/2$. The EoS parameter for the dilaton scalar
field is given by
\begin{equation}
\omega_{D}=\frac{p_D}{\rho_D}=\frac{-1+c'e^{\lambda\phi}\chi}{-1+3c'e^{\lambda\phi}\chi}.\label{wd}
\end{equation}
Equating Eq. (\ref{wd}) with the new holographic EoS parameter
(\ref{wh2}), $\omega_{D}=\omega_{\Lambda}$, we find the following
solution


\begin{equation}
c'e^{\lambda\phi}\chi=\frac{1}{3}\left(\frac{2k\Big(\frac{\alpha-\beta}{\alpha-\beta-1}\Big)e^{-2x}+\gamma\Big(\frac{3\beta-\alpha+1}{\beta}\Big)e^{-\frac{2}{\beta}(\alpha-1)x}}
{k\Big(\frac{\alpha-\beta}{\alpha-\beta-1}\Big)e^{-2x}+\gamma\Big(\frac{2\beta-\alpha+1}{\beta}\Big)
e^{-\frac{2}{\beta}(\alpha-1)x}}\right),\label{xid2}
\end{equation}
then using $\chi=\dot{\phi}^2/2$, we obtain
\begin{equation}
e^{\frac{\lambda\phi}{2}}\dot{\phi}=\sqrt{\frac{2}{3c'}}\left(\frac{2k\Big(\frac{\alpha-\beta}{\alpha-\beta-1}\Big)e^{-2x}
+\gamma\Big(\frac{3\beta-\alpha+1}{\beta}\Big)e^{-\frac{2}{\beta}(\alpha-1)x}}
{k\Big(\frac{\alpha-\beta}{\alpha-\beta-1}\Big)e^{-2x}+\gamma\Big(\frac{2\beta-\alpha+1}{\beta}\Big)e^{-\frac{2}{\beta}
(\alpha-1)x}}\right)^{1/2}.\label{phid1}
\end{equation}
Using $\dot{\phi}=\phi' H$ and integrating with respect to $x$ we
get
\begin{equation}
e^{\frac{\lambda\phi(a)}{2}}=e^{\frac{\lambda\phi(0)}{2}}+\frac{\lambda}{\sqrt{6c'}}\int_{0}^{\ln
a}\frac{1}{H}\left(\frac{2k\Big(\frac{\alpha-\beta}{\alpha-\beta-1}\Big)e^{-2x}+\gamma\Big(\frac{3\beta-\alpha+1}{\beta}\Big)e^{-\frac{2}{\beta}(\alpha-1)x}}
{k\Big(\frac{\alpha-\beta}{\alpha-\beta-1}\Big)e^{-2x}+\gamma\Big(\frac{2\beta-\alpha+1}{\beta}\Big)e^{-\frac{2}{\beta}
(\alpha-1)x}}\right)^{1/2}{\rm d}x,
\end{equation}
where $H$ is given by Eq. (\ref{H2-1}). Finally the evolutionary
form of the dilaton scalar filed is written as

\begin{equation}
\phi(a)=\frac{2}{\lambda}\ln
\left[e^{\frac{\lambda\phi(0)}{2}}+\frac{\lambda}{\sqrt{6c'}}\int_{0}^{\ln
a}\frac{1}{H}\left(\frac{2k\Big(\frac{\alpha-\beta}{\alpha-\beta-1}\Big)e^{-2x}+\gamma\Big(\frac{3\beta-\alpha+1}{\beta}\Big)e^{-\frac{2}{\beta}(\alpha-1)x}}
{k\Big(\frac{\alpha-\beta}{\alpha-\beta-1}\Big)e^{-2x}+\gamma\Big(\frac{2\beta-\alpha+1}{\beta}\Big)e^{-\frac{2}{\beta}
(\alpha-1)x}}\right)^{1/2}{\rm d}x\right].\label{phid2}
\end{equation}
For the flat case if we set $k=0$, using $\dot{x}=H$, assuming
$\phi(0)=-\infty$ for the present time $t_0=0$, then Eq.
(\ref{phid2}) reduces to
\begin{equation}
\phi(t)=\frac{2}{\lambda}\ln
\left[\frac{\lambda}{\sqrt{6c'}}\left(\frac{3\beta-\alpha+1}{2\beta-\alpha+1}\right)^{1/2}t\right],\label{phid}
\end{equation}
which is same as Eq. (3.24) in \cite{Granda2}.

\section{Conclusions}
We used a holographic DE model with new infrared cut-off
introduced by Granda and Oliveros \cite{Granda1}, which includes a
term proportional to $\dot{H}$. Contrary to the holographic DE
based on the event horizon, this model depends on local
quantities, avoiding in this way the causality problem. Hence the
proposed new infrared cut-off can be considered as a viable
phenomenological model of holographic density. We extended the
work of Granda and Oliveros \cite{Granda2} to the non-flat case.
However, some experimental data have implied that our universe is
not a perfectly flat universe and that it possesses a small
positive curvature ($\Omega_{k}\sim0.02$) \cite{Bennett}. Although
it is believed that our universe is flat, a contribution to the
Friedmann equation from spatial curvature is still possible if the
number of e-foldings is not very large \cite{Huang}. We obtained
the EoS parameter of the new holographic DE in the context of the
non-flat universe which, contrary to the flat case, is
time-dependent. This allows the model to transit from
$\omega_{\Lambda}>-1$ to $\omega_{\Lambda}<-1$ as indicated by
recent observations \cite{Alam}. We established a correspondence
between the new holographic DE energy density with the
quintessence, tachyon, K-essence and dilaton energy density in the
non-flat FRW universe. We reconstructed the potentials and the
dynamics of these scalar field models which describe quintessence,
tachyon, K-essence and dilaton cosmology. In the limiting case of
a flat universe, the results are in exact agreement with those
obtained by Granda and Oliveros \cite{Granda2}.


\end{document}